\documentclass[twocolumn,superscriptaddress,groupedaddress,amsmath,amssymb,aps,pre,floatfix,longbibliography]{revtex4-2}

\usepackage{comment}
\usepackage{graphicx}
\usepackage{dcolumn}
\usepackage{bm}

\usepackage[utf8]{inputenc}
\usepackage[T1]{fontenc}

\usepackage{braket}
\usepackage{siunitx}
\usepackage{xcolor}

\usepackage[normalem]{ulem}

\begin{document}

\title{Solitonic Solutions of the One-Dimensional Harmonically Trapped Repulsive Bose–Einstein Condensate via Neural Network Quantum States}
\author{Gaoqing Meng}
\affiliation{ 
Hebei Key Laboratory of Physics and Energy Technology, Department of Mathematics and Physics, North China Electric Power University, Baoding, Hebei 071003, China
}
\author{Mingshu Zhao}
\email{zmshum@ncepu.edu.cn}
\affiliation{ 
Hebei Key Laboratory of Physics and Energy Technology, Department of Mathematics and Physics, North China Electric Power University, Baoding, Hebei 071003, China
}

\date{\today} 

\begin{abstract}
We demonstrate the existence of bright solitons in a repulsively interacting, harmonically trapped quasi-one-dimensional Bose–Einstein condensate described by the Gross–Pitaevskii equation. Using a neural-network quantum state (NNQS) approach, we parametrize the initial wavefunction and optimize it to find solutions that recur after one trap period, effectively balancing repulsion with trap-induced attraction. 
Aside from the bright solitonic solution, we also report double bright and dark soliton states. 
Perturbing the initial state with multiplicative phase and amplitude noise confirms that these periodic orbits are orbitally stable.
Our results indicate that NNQS provides a powerful framework for uncovering coherent structures in nonlinear wave systems.
\end{abstract}

\maketitle

\section{Introduction}

Solitons are particle-like wave structures that emerge from the balance between nonlinearity and dispersion (or diffraction) in a medium, giving rise to localized, self-sustaining wave packets that preserve their shape during propagation. Solitons have been extensively studied across diverse physical settings, including optical fibers~\cite{haus1996solitons,song2019recent,sun2024multimode}, plasmas~\cite{kuznetsov1986soliton,ruderman2002propagation}, and Bose–Einstein condensates (BECs)~\cite{kengne2021spatiotemporal,mihalache2024localized} .
Different nonlinear models, such as the Korteweg–de Vries, sine-Gordon, and nonlinear Schr\"odinger (NLS) equations, capture distinct aspects of soliton behavior in these systems~\cite{kivshar1989dynamics}, yet the underlying principle of balancing opposing physical effects remains universal. 
Depending on the system parameters and the nature of the nonlinearity, this balance can manifest in a rich variety of coherent structures, including bright and dark solitons, breathers, and vortex solitons~\cite{kevrekidis2016solitons,flach2008,malomed2019vortex}, while extreme modulation instability can further give rise to rogue waves~\cite{dudley2014instabilities}.

The NLS equation describes the propagation of various types of solitons in nonlinear dispersive media. 
In fiber optics and photonic structures, bright solitons appear as localized intensity peaks---self-confined pulses of light---propagating on a zero background~\cite{hasegawa1973transmission}, while dark solitons manifest as localized dips, or notches of reduced intensity, embedded within a continuous-wave background~\cite{emplit1987picosecond,kivshar1993dark}. 
This framework extends to atomic matter-wave solitons in BECs, where the mean-field Gross–Pitaevskii equation (GPE) takes the form of an NLS-type equation, with atomic interactions providing the nonlinearity~\cite{kh2005dynamics,frantzeskakis2010dark}. In this system, the character of nonlinear matter-wave excitations is determined predominantly by the sign of atomic interactions and the external potential: repulsively interacting BECs support dark solitons~\cite{burger1999dark,denschlag2000generating,busch2000motion}, localized density dips with a characteristic phase slip, which can form periodic trains~\cite{strecker2002formation}. In contrast, attractively interacting BECs host robust, particle-like bright solitons~\cite{khaykovich2002formation,cornish2006formation}. Periodically modulated potentials, such as optical lattices, further enable gap solitons in repulsive systems~\cite{eiermann2004bright}.

In harmonically trapped BECs, the confining potential breaks translational invariance, rendering the system non-integrable. Consequently, the analytic solitonic solutions found in uniform media are replaced by oscillating solitary waves or breathers. While oscillating dark solitary waves in repulsive condensates are well-documented~\cite{burger1999dark, busch2000motion}, analogous bright solitary waves---localized density peaks sustained by a balance between repulsion and confinement---have not been theoretically established. However, Kohn's theorem dictates that the center-of-mass (COM) motion of any localized density wave packet must oscillate with the trap period $T_{\rm trap} = 2\pi/\omega$, where $\omega$ is the angular frequency of the trap~\cite{kohn1961cyclotron, brey1989optical, dobson1994harmonic}. A bright solitary wave would therefore constitute a nonlinear, non-diffracting dipole mode. Although the non-integrable dynamics can lead to chaotic evolution~\cite{Zhao2025Chaos}, the system's quasi-integrability~\cite{bland2018probing} suggests that regular, periodic trajectories may exist. Physically, the harmonic potential provides an effective attractive force that can counterbalance both dispersion and repulsive interactions. This presents a compelling, yet unresolved, theoretical possibility: the existence of bright solitary waves in a repulsive, harmonically trapped BEC.

Motivated by this possibility, the search for such a periodic solution can be framed as an optimization problem: minimizing the difference between an initial state and the state after evolution through one trap period $T_{\rm trap}$. 
To solve this, we employ the framework of neural-network quantum state (NNQS), which has emerged as a powerful paradigm for representing wavefunctions in quantum many-body systems, from lattice models to continuous spaces~\cite{medvidovic2024neural,lange2024architectures}. 
Originally developed for linear Schr\"odinger equations --- where it parameterizes wavefunctions to minimize energy functionals for ground states or to model time-dependent dynamics~\cite{gutierrez2022real} --- the method has recently been extended to nonlinear systems such as the GPE for computing ground and excited states~\cite{bao2025computing}.
In our work, we parameterize the initial wavefunction  $\Psi_0$ as an NNQS. We then time-evolve it under the GPE for one full period $T_0$ to obtain $\Psi_T$. 
By optimizing the network parameters to minimize the density difference between $\Psi_0$ and $\Psi_T$, we directly target a recurring solitonic solution, identified when this difference approaches zero.

Our numerical approach combines two complementary techniques: traditional high‑precision time integration (e.g., time‑splitting spectral schemes) and neural network methods.
Traditional high-precision methods are well‑suited for deterministic time evolution but are not inherently designed for inverse problems like ours; they lack a straightforward gradient-based optimization pathway back to the initial conditions. 
Conversely, neural network methods, such as NNQS and Physics-Informed Neural Networks~\cite{RAISSI2019686}, are built for optimization and can solve inverse problems, but they often struggle with the precision and long-time stability required for highly nonlinear dynamical systems. 
We combine the advantages of both: we employ a classical time-integration method to ensure accurate forward propagation of the NNQS, while using the auto-differentiation framework of the neural network to compute gradients of the final state with respect to the initial state parameters. 
This enables efficient backward propagation and gradient-based optimization of the initial wavefunction, which would be difficult to implement in a purely traditional numerical setting.

In this work, we solve the bright and dark solitonic solution in repulsive 1D harmonic trapped GPE using NNQS approach. To our best knowledge, this is the first report of bright solitons in repulsive harmonic trapped BECs. We also obtain the double solitonic solutions. 
We finally analyze the stability of the obtained solitons. The remainder of this work is structured as follows.
We first introduce the general property of the harmonic trapped GPE in section~\ref{sec:HT}. 
We then introduce the numerical schemes in section~\ref{sec:Numerical}. 
We later discuss the obtained solitonic solutions in section~\ref{sec:soliton}. We finally analyze the stability in section~\ref{sec:stability}.

\section{Harmonically Trapped BECs}\label{sec:HT}
\subsection{Gross-Pitaevskii Equation}
The behavior of a zero-temperature, weakly interacting BEC confined in a one-dimensional harmonic potential is captured by the time-dependent GPE under mean field approximation with the Hamiltonian $H[\psi]$:
\begin{equation}
    H[\psi]=-\frac{\hbar^2}{2m}\partial^2_x+\frac12m\omega^2 x^2+g_{\rm 1D} |\psi(x,t)|^2,
\label{eq:H_dim}
\end{equation}
\begin{equation}
i\hbar \frac{\partial {\psi}({x}, {t})}{\partial {t}} = 
 H[\psi]\ {\psi}({x},{t}),
\label{eq:GPE_dim}
\end{equation}
where ${\psi}({x},{t})$ representing the macroscopic wavefunction. 
The spatial and temporal variables are ${x}$ and ${t}$, respectively, $m$ denotes the atomic mass, $\omega$ the angular frequency of the confining trap, and ${g}_{\text{1D}}$ the strength of the interatomic interactions. 

A normalization condition is imposed on the wavefunction:
\begin{equation}
\int_{-\infty}^{\infty} |{\psi}({x},{t})|^2  d{x} = N,
\end{equation}
where $N$ is the total atom number.

We transition to a dimensionless framework via the scaling transformations:
\begin{equation}
X = {x} \sqrt{\frac{m \omega}{\hbar}}, \
T = \omega {t}, \
\Psi(X, T) = {\psi}({x}, {t}) \left(\frac{m \omega N^2}{\hbar}\right)^{-\frac14}.
\end{equation}
Here, $X$, $T$, and $\Psi(X, T)$ are the corresponding dimensionless variables. This rescaling yields a dimensionless coupling constant $g$, defined as
\begin{equation}
g = \frac{{g}_{\text{1D}} Nm^{1/2}}{\hbar^{3/2}\omega^{1/2}}.
\end{equation}

In these scaled units, the GPE adopts the form:
\begin{equation}
i \frac{\partial \Psi(X, T)}{\partial T} = 
\left( -\frac{1}{2} \frac{\partial^2}{\partial X^2} + \frac{1}{2} X^2 + g |\Psi(X, T)|^2 \right) \Psi(X, T),
\label{eq:GPE_dimless}
\end{equation}
subject to the normalization constraint:
\begin{equation}
\int_{-\infty}^{\infty} |\Psi(X, T)|^2  dX = 1.
\end{equation}
Throughout this work, we employ a value of $g = 100$ for the interaction parameter. For a $^{87}\mathrm{Rb}$ BEC in a trap with angular frequency $\omega=2\pi\times20\ \mathrm{Hz}$, this parameter choice yields a speed of sound $c \approx 1.6\ \mathrm{mm/s}$, a value that falls within the typical range observed in experiments~\cite{zhao2025kolmogorov, andrews1997propagation}.

\subsection{Kohn Theorem}
We use the dimensional form of the GPE in the remainder of this section to clearly show the effect of the external trap.

The Hamiltonian of the Eq.~\eqref{eq:H_dim} in the operator form can be written as 
\begin{equation}
    \hat H[\psi]=\frac{\hat p^2}{2m}+\frac12m\omega^2\hat x^2+g_{\rm 1D} |\psi(x,t)|^2,
\end{equation}
where $\hat p$ is the single-particle momentum operator.
We use the Heisenberg equation of motion
\begin{equation}
    \frac{d}{dt}\langle \hat{O} \rangle = \frac{i}{\hbar} \langle [\hat{H}(\psi), \hat{O}] \rangle
\end{equation}
to obtain the equation of motion for the COM. For $\hat{O} = \hat{x}$, we have
\begin{equation}
    \frac{d}{dt}\langle \hat{x} \rangle = \frac{i}{\hbar} \langle [\hat{H}(\psi), \hat{x}] \rangle = \frac{\langle \hat{p} \rangle}{m},
    \label{eq:EOM x}
\end{equation}
The equation of motion for $\langle\hat p\rangle$ is
\begin{equation}
    \frac{d}{dt}\langle \hat p\rangle=\frac{i}{\hbar}\langle[\hat H[\psi],\hat p]\rangle=-m\omega^2\langle\hat x\rangle,
    \label{eq:EOM p}
\end{equation}
where the interaction contribution is 0 since
\begin{align}
    &\langle[\hat p,|\psi(x,t)|^2]\rangle=-i\hbar\langle\partial_x|\psi(x,t)|^2\rangle=0.
\end{align}
Combining Eq.\eqref{eq:EOM x} and Eq.~\eqref{eq:EOM p} we obtain 
\begin{equation}
    \frac{d^2}{dt^2}\langle\hat x\rangle=-\omega^2\langle\hat x\rangle.
\end{equation}



This establishes the Kohn theorem: the COM obeys the differential equation for simple harmonic motion with an angular frequency $\omega$. 
Consequently, the density profile of the solitonic solution in dimensionless units, $|\Psi(X,T)|^2$, is periodic in time with period $T_0 = 2\pi$.

It is important to note that the above derivation assumes the wavefunction vanish at the boundaries: $\psi \to 0$ as $x \to \pm\infty$.

\subsection{Harmonic trap squeezing of the wave packet}
In this subsection, we derive the equation of motion for the square of the width, $\sigma_x^2 = \langle \hat x^2 \rangle - \langle \hat x \rangle^2$, in a harmonic trap.
We calculate the first order time derivative of $\langle\hat x\rangle^2$ using Eq.~\eqref{eq:EOM x}
\begin{equation}
    \frac{d}{dt}\langle \hat x\rangle^2=2\langle\hat x\rangle\frac{d\langle\hat{x}\rangle}{dt}=\frac{2\langle\hat x\rangle\langle\hat p\rangle}{m}.
\end{equation}
We then calculate the second order time derivative using Eq.~\eqref{eq:EOM x} and Eq.~\eqref{eq:EOM p}
\begin{equation}
    \frac{d^2}{dt^2}\langle\hat x\rangle^2=\frac{2}{m}\left(\frac{d\langle\hat x\rangle}{dt}\langle \hat p\rangle+\langle\hat x\rangle\frac{d\langle\hat p\rangle}{dt}\right)=\frac{2\langle\hat p\rangle^2}{m^2}-2\omega^2\langle \hat x\rangle^2.
\end{equation}

Similarly, we use the Heisenberg equation of motion to calculate the first order derivative of $\langle \hat x^2\rangle$
\begin{equation}
    \frac{d\langle\hat x^2\rangle}{dt}=\frac{i}{\hbar}\langle[\hat H,\hat x^2]\rangle=\frac{i}{2m\hbar}\langle[\hat p^2,\hat x^2]\rangle=\frac{\langle\hat x\hat p+\hat p\hat x\rangle}{m}.
\end{equation}
Next, we apply the Heisenberg equation of motion again to obtain the second derivative:
\begin{equation}
    \frac{d^2}{dt^2}\langle\hat x^2\rangle=\frac{i}{m\hbar}\langle[\hat H,\hat x\hat p+\hat p\hat x]\rangle
\end{equation}
We evaluate the commutator separately for the kinetic, potential and interaction parts.
For the kinetic term:
\begin{equation}
    \langle[\frac{\hat p^2}{2m},\hat x\hat p]\rangle+\langle[\frac{\hat p^2}{2m},\hat p\hat x]\rangle=\frac{-2i\hbar}{m}\langle\hat p^2\rangle.
\end{equation}
For the harmonic potential term:
\begin{equation}
    \langle[\frac12m\omega^2\hat x^2,\hat x\hat p]\rangle+\langle[\frac12m\omega^2\hat x^2,\hat p\hat x]\rangle=2i\hbar m\omega^2\langle\hat x^2\rangle.
\end{equation}
For the interaction term:
\begin{align}
   &\langle[g_{\rm 1D}|\psi|^2,\hat x\hat p]\rangle+\langle[g_{\rm 1D}|\psi|^2,\hat p\hat x]\rangle
  =-i\hbar g_{\rm 1D}\int dx\ \rho^2(x,t),
\end{align}
where $\rho(x,t)=|\psi(x,t)|^2$.

Combining above results, we obtain
\begin{equation}
    \frac{d^2}{dt^2}\langle\hat x^2\rangle=\frac{2\langle\hat p^2\rangle}{m^2}-2\omega^2\langle\hat x^2\rangle+\frac{g_{\rm 1D}}{m}\int dx\ \rho^2(x,t).
\end{equation}

We finally obtain the equation of motion for the $\sigma_x^2$
\begin{equation}
    \frac{d^2}{dt^2}(\sigma_x^2)=\frac{2\sigma_p^2}{m^2}-2\omega^2\sigma_x^2+\frac{g_{\rm 1D}}{m}\int dx\ \rho^2(x,t),
\end{equation}
where $\sigma_p^2\equiv \langle\hat p^2\rangle-\langle\hat p\rangle^2$.
We can identify three distinct contributions: the first term, ${2\sigma_p^2}/{m^2}$, arises from quantum pressure (i.e., kinetic spreading); the second term, $-2\omega^2\sigma_x^2$, is due to the harmonic trapping; and the third term, ${g_{\rm 1D}}\int \rho^2\,dx/m$, originates from repulsive interatomic interactions. The harmonic trapping term is negative and therefore counteracts the other two effects, which both tend to increase the width squared.
For example, consider the case that the wave packet initially has zero velocity, and the $\frac{d}{dt}(\sigma_x^2)=0$, then the quantum pressure and the repulsive interaction induces the diffusion effect since they gives a positive width spreading velocity, while the trapping term counteracts. And these three effects might be able to balance to gives a periodic orbit.
A special case would be the noninteracting case, when $g_{\rm 1D}=0$, the steady state with constant $\sigma_x^2$ exists for the coherent state. With $g_{\rm 1D}>0$, the width is oscillating during a period, so the solitons in the harmonic trap should be more precisely defined as the solitary wave.

\section{Numerical Methods}\label{sec:Numerical}

\subsection{NNQS}
Our goal is to obtain the soliton solutions of the dimensionless GPE given in Eq.~\eqref{eq:GPE_dimless}.
To this end, we employ a NNQS framework to represent the initial dimensionless wavefunction $\Psi(X, T=0)$.
A key simplification arises from our focus on initial states with zero initial velocity, which allows the initial wavefunction to be represented by a real-valued function, thereby reducing the complexity of the neural network representation. The network architecture consists of a fully-connected multilayer perceptron (MLP)~\cite{rumelhart1986learning} that maps the spatial coordinate $X$ to the corresponding wavefunction value $\Psi(X)$. 
We employ the Gaussian error linear unit (GeLU) activation function~\cite{hendrycks2016gaussian}, which produces a smooth and physically realistic wavefunction profile.
The detailed NNQS architecture is summarized in Table~\ref{tab:NNQS}, demonstrating a design that provides sufficient expressivity to capture complex profiles while maintaining computational efficiency.
Our MLP model is implemented using PyTorch~\cite{paszke2019pytorch}.
\begin{table}[bth]
\centering
\begin{tabular}{l}
\hline
Input: $X$ (size 1) \\
\quad Linear(1, 256) $\rightarrow$ GeLU \\
\quad Linear(256, 256) $\rightarrow$ GeLU \\
\quad Linear(256, 128) $\rightarrow$ GeLU \\
\quad Linear(128, 64) $\rightarrow$ GeLU \\
\quad Linear(64, 1) \\
Output: $\Psi(X)$ (size 1) \\
\hline
\end{tabular}
\vspace{0.15cm}
\caption{\label{tab:NNQS}%
Network architecture for the 1D NNQS framework.}
\end{table}

\subsection{Time evolution of NNQS}
The time evolution of the wavefunction under the GPE is implemented using a time-splitting spectral method~\cite{bao2003numerical}. For a single period $T_0 = 2\pi$, we discretize the evolution into $N_{T} = 40,000$ time steps with $\Delta T \approx 1.57 \times 10^{-4}$. The evolution operator is split into three parts:
\begin{equation}
\bar\Psi(X,T+\frac{\Delta T}{2}) = e^{-i(V_{\text{ext}} + g|\Psi|^2)\Delta T/2}\Psi(X,T),
\end{equation}
\begin{equation}
\widetilde\Psi = \mathcal{F}^{-1}\left[e^{-i k^2 \Delta T/2} \mathcal{F}[\bar\Psi(X,T+\frac{\Delta T}{2})]\right],
\end{equation}
\begin{equation}
    \Psi(X,T+\Delta T)=e^{-i(V_{\text{ext}} + g|\widetilde\Psi|^2)\Delta T/2} \widetilde\Psi,
\end{equation}
where the potential $V_{\rm ext}=\frac12X^2$ and nonlinear terms are handled in real space using exponential operators; the kinetic energy operator is evaluated in Fourier space using fast Fourier transforms denoted by $\mathcal{F}$; wavefunction normalization is enforced at each time step to maintain $\int |\Psi|^2 dX = 1$.
The spatial domain is discretized over $X \in [-32, 32]$ with $N_{X} = 4097$ grid points.

The composite time evolution over one period, represented by the operator $\mathcal{U}_T$ such that $\Psi(X,T) = \mathcal{U}_T\Psi(X,0)$, is fully differentiable with respect to the initial wavefunction $\Psi(X,0)$ produced by the neural network. Because every operation in the evolution—including the time-splitting spectral method, pointwise nonlinearities, and normalization—is implemented using differentiable PyTorch primitives, gradients can be obtained via automatic differentiation~\cite{baydin2018automatic}. 
This enables end-to-end training of the NNQS.

\subsection{Loss Functionals and optimizations}
The loss functional is defined via the density difference $\rm L_1$ norm in addition to a penalty constraints on the initial COM.
The loss functional combines multiple physical constraints:
\begin{equation}
\mathcal{L} = \lambda_{\rm residual} \mathcal{L}_{\text{residual}} + \lambda_{\rm COM} \mathcal{L}_{\text{COM}}, 
\end{equation}
where $\mathcal{L}_{\text{residual}} = \int \big||\Psi(X,T)|^2 - |\Psi(X,0)|^2 \big| dX$ ensures periodicity of the density profile. $\mathcal{L}_{\text{COM}} = (\langle X\rangle_{t=0} - \bar{X}_0)^2$ constrains the initial COM position at $\langle X\rangle_{t=0}=\bar{X}_0$.
The coefficients $\lambda_{\rm residual}$ and $\lambda_{\rm COM}$ are penalty strengths that balance the relative importance of their respective loss terms.
We can add more constraints given our requirement of the solutions, which will be elaborated in section~\ref{sec:soliton}.

The optimization is implemented with Adam optimizer~\cite{kingma2014adam}.
The Adam optimizer integrates the benefits of two key concepts: momentum, which accelerates convergence by maintaining a moving average of past gradients, and RMSProp~\cite{tieleman2012lecture}, which adapts the learning rate $\eta$ by scaling it according to a moving average of squared gradients. This combination facilitates efficient and stable optimization.

For a parameter $\theta_i$ at optimization step $\bar t$, the update rule is defined as:
\begin{equation}
\theta_i^{(\bar t+1)} = \theta_i^{(\bar t)} - \eta \frac{\hat{m}_i^{(\bar t)}}{\sqrt{\hat{v}_i^{(\bar t)}} + \kappa}.
\end{equation}
Here, $\hat{m}_i$ and $\hat{v}_i$ represent the bias-corrected estimates of the first moment (mean) and second moment (uncentered variance) of the gradients for $\theta_i$, respectively. The term $\kappa$ (set to $10^{-8}$) is a small constant that ensures numerical stability by preventing division by zero. The exponential decay rates for the moment estimates are configured with standard values of $\beta_1 = 0.9$ and $\beta_2 = 0.999$.

\section{Solitonic Solutions}\label{sec:soliton}
\subsection{Single Bright Soliton}
\begin{figure}[bth]
\centering
\includegraphics{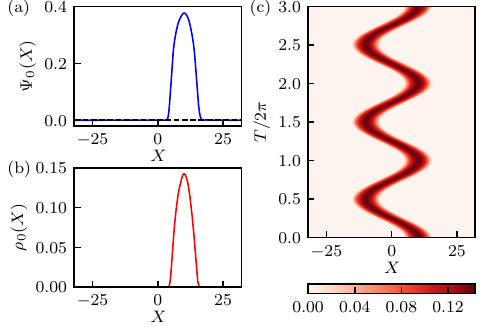}
\caption{Single bright soliton in harmonic trap.(a) NNQS solution of the initial state $\Psi_0(X)$ with COM at $\bar{X}_0=10$. The black dashed line marks $\Psi_0(X)=0$.(b) Initial density profile $\rho_0(X)$.
(c) Density profile evolution $\rho({X,T})$.}
\label{fig:single_bright_soliton}
\end{figure}

The bright soliton solution is a localized state; therefore, we employ the position-shifted ground state solution in the harmonic trap as the initial guess for the NNQS optimization. 
The penalty strengths are chosen as $\lambda_{\rm residual}=10^4$ and $\lambda_{\rm COM}=10^2$.
The optimization is performed using the Adam optimizer with an exponential decay learning rate from $10^{-5}$ to $10^{-6}$ for approximately $4000$ epochs, reducing the residual loss $\mathcal{L}_{\rm residual}$ below $4\times10^{-3}$. 
The initial COM position is constrained to $\bar{X}_0 = 10$. The resulting wavefunction $\Psi_0(X)$ and density profile $\rho_0(X)$ are presented in Fig.~\ref{fig:single_bright_soliton}(a-b). These profiles are slightly asymmetric about the COM with skewness $1.4\times10^{-3}$ due to the influence of the harmonic trapping potential. To verify the bright soliton nature of the solution, we perform time evolution using a time-splitting spectral method for three harmonic cycles. The corresponding density profile trajectory $\rho(X,T)$, shown in Fig.~\ref{fig:single_bright_soliton}(c), demonstrates a periodic evolution while maintaining a localized density profile greater than zero, confirming it as a bright soliton.

It is worth noting that the existence of bright soliton solution is independent of the initial COM position $\bar X_0$. 
We selected a relatively large value of $\bar X_0 = 10$ specifically to analyze the regime where the dipole-mode approximation is invalid. The NNQS approach proves robust, yielding stable solutions even when using less smooth activation functions like rectified linear unit (ReLU)~\cite{fukushima2007visual}. Empirically, we find that a residual loss $\mathcal{L}_{\rm residual}$ below $10^{-2}$ is typically sufficient for obtaining accurate results. A more detailed stability analysis involving multiplicative noise is presented in the section~\ref{sec:stability}.

\subsection{Single Dark Soliton}
\begin{figure}[htb]
\centering
\includegraphics{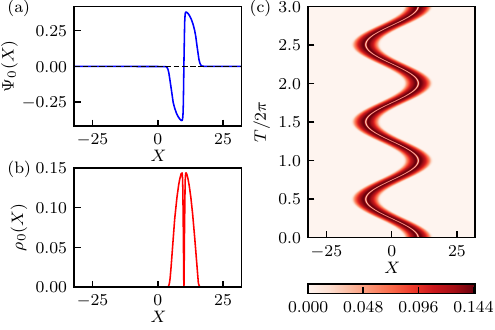}
\caption{Single dark soliton in harmonic trap.
(a) NNQS solution of the initial state $\Psi_0(X)$ with COM at $\bar{X}_0=10$. The black dashed line marks $\Psi_0(X)=0$.(b) Initial density profile $\rho_0(X)$.
(c) Density profile evolution $\rho({X,T})$.}
\label{fig:single_dark_soliton}
\end{figure}

We discuss the result of single dark soliton solution using NNQS optimization in this subsection. As a dark soliton is characterized by a $\pi$ phase jump, it is topologically analogous to the first excited state of the harmonic trap. 
We therefore use this excited state, shifted to the constrained initial COM position $\bar{X}_0 = 10$, as the initial guess. 
The optimization parameters (e.g., optimizer, learning rate, epoch count) are identical to those used for the single bright soliton. 
The resulting initial wavefunction $\Psi_0(X)$ and density $\rho_0(X)$ are shown in Fig.~\ref{fig:single_dark_soliton}(a-b). The harmonic confinement induces a clear asymmetry in the profile relative to the density minimum. The solution is verified by time evolution over three harmonic cycles; the resulting dynamics $\rho(X,T)$ are presented in Fig.~\ref{fig:single_dark_soliton}(c). 
While the overall envelope oscillates similarly to the bright soliton, the sustained density dip at its center confirms the dark soliton character of the solution.

It is important to note that the dark soliton realized in our system differs from the conventional one often discussed in the literature~\cite{burger1999dark}. 
In the traditional picture, a dark soliton can be viewed as a coherent mixture of the ground state and the first excited state of the harmonic trap. For strong interactions ($g\gg1$), the resulting density dip oscillates at a frequency close to $\omega/\sqrt{2}$~\cite{becker2008oscillations}, which is distinct from the Kohn mode frequency $\omega$ appearing in our settings. 
Moreover, such a mixture is generally not stable over long times and tends to exhibit chaotic dynamics~\cite{Zhao2025Chaos}. 
In contrast, the dark soliton presented here propagates together with a bright component of the localized wave packet, so that the density dip moves within a bright background---unlike the traditional case where the dip oscillates in an otherwise nearly stationary condensate density background. 
This key structural difference underlies the distinct dynamical behavior observed in our system.

\subsection{Imbalanced Double Bright Soliton}


In integrable systems, the existence of a single-soliton solution typically implies the existence of multi-soliton states. While harmonic trapping breaks exact integrability, the system is often treated as quasi-integrable; consequently, stable multi-soliton solutions are more likely to persist than in strongly non-integrable settings. 

To guide the optimizer away from converging to a single-soliton solution, we introduce additional constraints. Two primary strategies can be employed: (i) penalizing the wavefunction's overlap with the known single-soliton solution, and (ii) incorporating an entropy-based loss term, $S = -\int dX \, \rho(X) \ln[\rho(X)]$, which applies a penalty if the entropy falls below that of the single-soliton state. In this work, we adopt the second method.
The single bright soliton has entropy $2.26$; therefore, a state consisting of two bright solitons is expected to have a larger entropy.
We set the target entropy to $S_2 = 2.55$ and define the entropy loss as $\mathcal{L}_{\rm entropy} = (S - S_2)^2$ with a penalty strength $\lambda_{\rm entropy} = 10^2$.

Furthermore, to mitigate artifacts from the built-in periodic boundary conditions of the spectral method when the wavefunction is not tightly localized, we add a Dirichlet boundary loss, $\mathcal{L}_{\mathrm{BC}} = |\Psi(X_0)| + |\Psi(X_{N_X-1})|$, which encourages the wavefunction to vanish at the computational domain edges.
The peanlty strength are set to $\lambda_{\rm BC}=10^3$.

\begin{figure}[bt]
\centering
\includegraphics{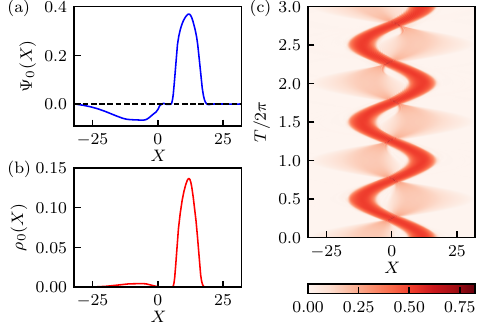}
\caption{Imbalanced double bright soliton in harmonic trap.(a) NNQS solution of the initial state $\Psi_0(X)$ with COM at $\bar{X}_0=10$. The black dashed line marks $\Psi_0(X)=0$.(b) Initial density profile $\rho_0(X)$.
(c) Transformed density profile evolution $\rho^{1/3}({X,T})$.}
\label{fig:double_bright_soliton}
\end{figure}

To find a double bright soliton solution, we use the single bright soliton state as the initial guess for the NNQS optimization. 
The initial COM position is again constrained to $\bar{X}_0 = 10$. 
The Adam optimizer is employed with a learning rate decaying exponentially from $10^{-6}$ to $10^{-7}$ over approximately $10000$ epochs, until the residual loss $\mathcal{L}_{\rm residual}$ falls below $10^{-2}$. 
The converged wavefunction $\Psi_0(X)$ and its density $\rho_0(X)$ are shown in Fig.~\ref{fig:double_bright_soliton}(a-b). The profile reveals two distinct localized peaks. 
Notably, the amplitude of the soliton on the negative $X$ side is $0.18$ times smaller than that of its positive-side counterpart. 
We verify the solution by time-evolving it for three harmonic cycles using the time-splitting spectral method. 
To enhance the visibility of the smaller-amplitude soliton in the density plot, we display the trajectory of the transformed density $\rho^{1/3}(X,T)$ in Fig.~\ref{fig:double_bright_soliton}(c). Both density peaks undergo coherent periodic motion while maintaining their localized, positive-definite character, confirming the state as a double bright soliton.

\subsection{Balanced Double bright Soliton}
In this subsection, we consider a balanced double bright soliton configuration, namely the collision of two identical bright solitons. Due to the symmetry of the setup---equal soliton amplitudes and opposite velocities during evolution---the COM remains fixed at $\bar{X}_0 = 0$.
To enforce an initial separation, we constrain the second moment of the density distribution to $\langle{(X-\bar{X}_0)^2}\rangle_{t=0} \equiv \int dX\, |\Psi_0(X)|^2 (X-\bar{X}_0)^2 = 105$, which positions each soliton approximately at $X \approx \pm 10$. 
Although the COM is stationary, the relative motion is periodic with the harmonic trap period $T_0 = 2\pi$.

The initial guess is constructed as the symmetric superposition of a single bright soliton profile: $\Psi_{\text{init}}(X) = \mathsf{N} \left[ \Psi_{\text{sb}}(X) + \Psi_{\text{sb}}(-X) \right]$, where $\mathsf{N}$ is a normalization factor and $\Psi_{\text{sb}}(X)$ is the single bright soliton solution in Fig.~\ref{fig:single_bright_soliton}(a). 
The loss function structure and optimization parameters (including the learning rate schedule and epoch count) are similar to those used for the imbalanced double soliton case, except that the entropy loss is not required.

\begin{figure}[htb]
\centering
\includegraphics{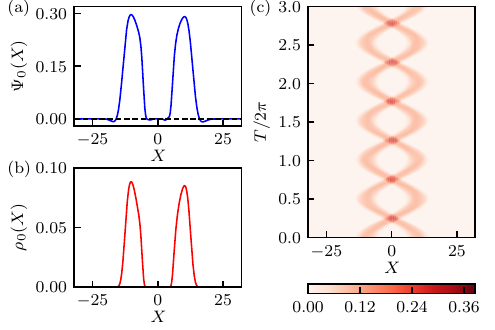}
\caption{Balanced double bright soliton in harmonic trap.(a) NNQS solution of the initial state $\Psi_0(X)$ with COM at $\bar{X}_0=10$. The black dashed line marks $\Psi_0(X)=0$.(b) Initial density profile $\rho_0(X)$.
(c) Density profile evolution $\rho({X,T})$.}
\label{fig:double_bright_soliton_collision}
\end{figure}

The converged initial wavefunction $\Psi_0(x)$ and its density $\rho_0(x)$ are presented in Fig.~\ref{fig:double_bright_soliton_collision}(a--b), showing two well-separated, nearly equal-amplitude peaks. We note that the wavefunctions are not perfectly symmetric. A symmetry loss could be introduced if perfect symmetry is required. During optimization, some ripples may appear near the boundaries, which can be eliminated by extending the Dirichlet boundary condition (zero-field condition) to the relevant regions.

We verify the solution by time-evolving the state for three harmonic cycles using the time-splitting spectral method. 
The resulting density trajectory $\rho(x,t)$ is shown in Fig.~\ref{fig:double_bright_soliton_collision}(c). 
The plot reveals coherent periodic motion of both density peaks, which maintain their localized, positive-definite character throughout the evolution---confirming the existence of a balanced double bright soliton solution. During each collision, a clear oscillatory interference pattern emerges.
This region is likely to be chaotic, as discussed in the section~\ref{sec:stability}.

\subsection{Balanced Double Dark Soliton}


We investigate the existence of double dark soliton in this subsection. 
For simplicity, we only consider the case of balanced double dark soliton with a constrained COM position of $\bar{X}_0 = 0$ and $\langle{(X-\bar{X}_0)^2}\rangle_{t=0}=105$ to ensure initial soliton spatial separations.

The initial guess is constructed as the symmetric superposition of a single dark soliton profile: $\Psi_{\text{init}}(X) = \mathsf{N} \left[ \Psi_{\text{sd}}(X) + \Psi_{\text{sd}}(-X) \right]$, where $\mathsf{N}$ is a normalization factor and $\Psi_{\text{sd}}(X)$ is the single dark soliton solution is Fig.~\ref{fig:single_dark_soliton}(a). 
The loss function structure and optimization parameters (including the learning rate schedule and epoch count) are the same to those used for the balanced double bright soliton case.

\begin{figure}[tbh]
\centering
\includegraphics{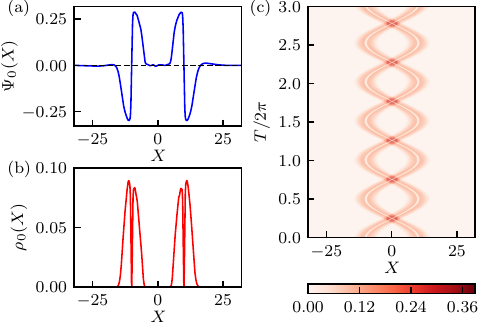}
\caption{Double dark soliton in harmonic trap.(a) NNQS solution of the initial state $\Psi_0(X)$ with COM at $\bar{X}_0=10$. The black dashed line marks $\Psi_0(X)=0$.(b) Initial density profile $\rho_0(X)$.
(c) Density profile evolution $\rho({X,T})$.}
\label{fig:double_dark_soliton}
\end{figure}

The initial wavefunction $\Psi_0(X)$ and the density $\rho_0(X)$ are shown in Fig.~\ref{fig:double_dark_soliton}(a--b). The dynamical stability of this state is verified by time evolution over three harmonic cycles. The trajectory of $\rho(X,T)$, presented in Fig.~\ref{fig:double_dark_soliton}(c), confirms the robust, periodic oscillation of the entire structure. The density dips associated with the dark solitons are clearly maintained throughout the evolution, demonstrating the existence of double dark soliton.

\section{Stability Analysis}\label{sec:stability}

We verify the stability by perturbing the initial state with multiplicative noise, which mimics the experimental noise present when using digital micromirror devices for phase and density imprinting~\cite{gauthier2016direct,zhao2025kolmogorov,fritsch2020creating}. The perturbed wave function is constructed as
\begin{equation}
\widetilde{\Psi}(X) = \Psi_0(X) \, e^{i \phi(X)} \, \sqrt{\zeta(X)},
\end{equation}
where $\phi(X) \sim \mathcal{N}(0,\sigma^2)$ and $\zeta(X) \sim \mathcal{N}(1,\epsilon^2)$ are independent Gaussian random fields. For the results presented here, we use $\sigma = 0.005\pi$ and $\epsilon = 0.0025$. The amplitude factor $\sqrt{\zeta(X)}$ introduces multiplicative noise in the density, while $e^{i\phi(X)}$ adds phase noise. After generating the perturbed state, we renormalize it to preserve the norm of the wavefunction.

\begin{figure}[tbh]
\centering
\includegraphics{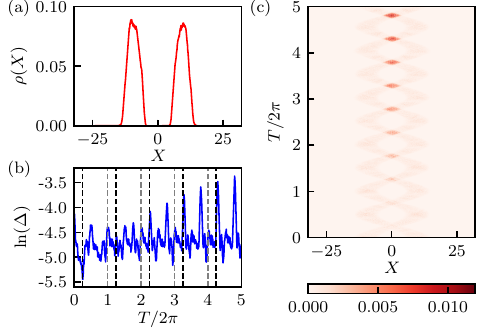}
\caption{{Stability of the balanced double bright soliton.(a) Perturbed balanced double bright soliton $\rho(X)$. (b) Distance $\Delta(T)$ versus time; grey dashed lines: $T=2n\pi$, black dashed lines: $T=(2n+\frac12)\pi$, where $n$ is an integer. (c) Density difference profile evolution $|\rho({X,T})-\rho_0({X,T})|$.}}
\label{fig:lyapunov}
\end{figure}

For each configuration we perform $10$ independent noise realizations. The evolution of the density difference is monitored using the distance
\begin{equation}
\Delta(T) = \int \bigl| \rho(X,T) - \rho_0(X,T) \bigr| \,dX,
\end{equation}
where $\rho(X,T)=|\widetilde\Psi(X,T)|^2$ is the perturbed density trajectory and $\rho_0(X,T)=|\Psi_0(X,T)|^2$ is the density of the unperturbed reference trajectory. An example of $\Delta(T)$ for the perturbed balanced double bright soliton is shown in Fig.~\ref{fig:lyapunov}(b). 
The distance $\Delta$ is large at times $T = 2n\pi$ (marked by grey dashed lines), when the two solitons are well separated, and becomes small during collisions (marked by black dashed lines), when the solitons approach each other.

To quantify the instability, we compute the Lyapunov exponent~\cite{strogatz2018nonlinear} from the slope of the logarithm of the distance versus time. We consider two distinct Poincaré sections, corresponding to stroboscopic sampling at two fixed phases of the periodic orbit. For each section, we evaluate
\begin{equation}
\lambda = \frac{d}{dT} \ln \Delta(T)
\end{equation}
by performing a linear fit to $\ln \Delta(T)$ over successive periods $T_0 = 2\pi$.
Specifically, we define two sequences of stroboscopic times:
$T_{k0} = 2n\pi$ (phase $0$) and $T_{k1} = (2n+\frac12)\pi$ (phase $\pi/2$), where $n$ is an integer.
We then fit $\ln \Delta(T_{k0})$ versus $T_{k0}$ to obtain $\lambda_0$, and separately fit $\ln \Delta(T_{k1})$ versus $T_{k1}$ to obtain $\lambda_1$.
The uncertainties for $\lambda_0$ and $\lambda_1$ are estimated as the standard error of the mean over the $10$ independent realizations.

The results for all soliton configurations considered in the previous section are summarized in Table~\ref{tab:Lyapunov}.
In every case we find $\lambda_0 < 0$ and $\lambda_1 > 0$. This indicates that the full periodic trajectory (the limit cycle) is stable in the sense of orbital stability, yet at the specific instants when the solitons acquire high kinetic energy, the dynamics exhibits transient instability.

\begin{table}[htb]
\begin{ruledtabular}
\begin{tabular}{l c c}
\textrm{Soliton configurations} & 
{$\lambda_0$} &{$\lambda_1$} 
 \\
\hline
single bright & $-6.8(6)\times10^{-3}$ &$1.3(1)\times10^{-2}$\\ 
single dark &  $-6(1)\times10^{-3}$ &$9(2)\times10^{-3}$\\
imbalanced double bright & $-2.9(7)\times10^{-3}$  &$2.1(1)\times10^{-2}$ \\
balanced double bright & $-5.1(4)\times10^{-3}$ &$2.3(1)\times10^{-2}$ \\ 
balanced double dark &  $-1.2(9)\times10^{-3}$ &$2.0(2)\times10^{-2}$ \\
\end{tabular}
\end{ruledtabular}
\caption{\label{tab:Lyapunov}%
Lyaunov exponent $\lambda$ for different soliton configurations.}
\end{table}

\section{Conclusion and Outlooks}
In this work, we have studied solitonic solutions of the one-dimensional GPE with a repulsive interaction and a harmonic trap. By parameterizing the initial wavefunction as a NNQS and minimizing the density difference between the initial state and the state obtained after one trap period $T_{\rm trap} = 2\pi/\omega$, we have identified periodic solitary-wave solutions.
Our hybrid approach combines accurate time-splitting spectral propagation with gradient-based optimization via automatic differentiation, enabling efficient end-to-end training. 
Using this method, we have obtained bright solitons, dark solitons, and double-soliton configurations in the repulsive, harmonically trapped condensate. 
To assess dynamical stability, we perturbed each soliton with multiplicative Gaussian noise in phase and amplitude and monitored the density distance between perturbed and unperturbed trajectories. 
The full periodic orbit is orbitally stable, although transient instabilities occur when soliton approaches trap center  and the kinetic energy is high. 
Our results demonstrate that NNQS combined with classical time evolution algorithm provide a powerful and flexible framework for discovering non-trivial solitonic and periodic solutions in nonlinear wave systems, opening new avenues for waveform design in trapped atomic condensates and beyond.

The neural-network-based optimization framework can be extended to study soliton--soliton interactions in greater detail, including collisions between multiple bright or dark solitons, as well as bright--dark soliton mixtures, where the coexistence of density dips and peaks in different components or internal states leads to rich nonlinear dynamics~\cite{kodama1987soliton}. 
Whether genuine higher-order localized states—such as triple or higher-order solitons—exist in this non-integrable, harmonically trapped system remains an open and challenging question; the present method offers a systematic way to search for them, but their existence is not guaranteed and would itself be a nontrivial discovery.  
For more complicated external potentials—such as double wells, optical lattices, or disordered potentials—the optimization framework remains applicable, but the Kohn theorem no longer holds, and the relevant return period for self-consistency is not known a priori. Determining both the period and the initial state simultaneously becomes a more challenging inverse problem, yet one that the neural-network approach is well-suited to address. 

Beyond one dimension, the method naturally generalizes to two- and three-dimensional trapped condensates, where vortex solitons, vortex rings, and other topological structures could be discovered as periodic solutions.
Furthermore, applying similar periodicity optimization to Floquet systems~\cite{bukov2015universal}, where the Hamiltonian is time-periodic (e.g., via modulated trapping or shaking), could reveal dynamically stabilized nonlinear modes. 
These extensions highlight the versatility of combining neural-network parameterization with gradient-based periodicity optimization for exploring exotic nonlinear wave phenomena.

\begin{acknowledgments}
This work was partially supported by the Research and Practice Project of Higher Education Teaching Reform of Hebei Province (No. 2025GJJG412, 2026GJJG415).
\end{acknowledgments}

\bibliography{main}

\end{document}